\documentclass[10pt, conference, letterpaper]{IEEEtran}

\ifCLASSINFOpdf
\else
\fi

\usepackage{algorithmic}
\usepackage{algorithm}
\usepackage{url}
\usepackage{multirow}
\usepackage{cite}
\usepackage[pdftex]{graphicx}

\begin{document}
\title{AccurateML: Information-aggregation-based Approximate Processing for Fast and Accurate Machine Learning on MapReduce}

\author{
\IEEEauthorblockN{Rui Han, Fan Zhang, Zhentao Wang}
\IEEEauthorblockA{Institute Of Computing Technology, Chinese Academy of Sciences\\
Beijing, China\\
\{hanrui, zhangfan, wangzhentao\}@ict.ac.cn
}
}

\maketitle

\begin{abstract}

The growing demands of processing massive datasets have promoted irresistible trends of running machine learning applications on MapReduce. When processing large input data, it is often of greater values to produce fast and accurate enough approximate results than slow exact results. Existing techniques produce approximate results by processing parts of the input data, thus incurring large accuracy losses when using short job execution times, because all the skipped input data potentially contributes to result accuracy.
We address this limitation by proposing AccurateML that aggregates information of input data in each map task to create small aggregated data points. These aggregated points enable all map tasks producing initial outputs quickly to save computation times and decrease the outputs' size to reduce communication times. Our approach further identifies the parts of input data most related to result accuracy, thus first using these parts to improve the produced outputs to minimize accuracy losses. We evaluated AccurateML using real machine learning applications and datasets. The results show: (i) it reduces execution times by 30 times with small accuracy losses compared to exact results; (ii) when using the same execution times, it achieves 2.71 times reductions in accuracy losses compared to existing approximate processing techniques.

\end{abstract}



\section{Introduction} \label{Section: Introduction}

A wide range of domains today such as signal processing, financial analysis, and multimedia processing require the running of machine learning (ML) applications on massive datasets to extract values and get crucial insights. This has led to an increasing interest in executing ML applications on MapReduce frameworks like Hadoop \cite{hadoop} and Spark \cite{sparkWebSite}.
Running a MapReduce job on a large-scale input data is usually a time-consuming process, in which the computation times of map tasks take a large proportion of the job execution time while the communication times in the shuffle phase (that is, transferring intermediate data between map and reduce tasks) take accounts for a third of this time in many applications (e.g. 68\% of applications in Facebook's Hadoop cluster \cite{chowdhury2011managing}).

Due to the high resource and time consumptions in processing entire input data with millions of data points to deliver exact results, a widely applied approach is to produce approximate results in order to trade off result accuracy (correctness) for job execution time reduction \cite{agarwal2013blinkdb,chippa2013analysis,goodEnoughComputing,han2017clap}.
For example, in k-nearest neighbors (kNN) classification \cite{cover1967nearest} and collaborative filtering (CF)-based recommendation \cite{su2009survey} applications, the \emph{result accuracies} are \emph{classification accuracies} and \emph{errors} between the predicted and the actual ratings, respectively.
As small accuracy losses cannot be evidently perceived and thus are tolerable by application users  \cite{goodEnoughComputing}, efficiently and successfully applying such approximate processing mechanism requires considerably reducing job execution time without incurring large losses in result accuracy.

Existing techniques reduce execution time of MapReduce jobs by restricting the size of the input data used to produce approximate results \cite{condie2010mapreduce, pansare2011online,li2011platform, laptev2012early, agarwal2014knowing}.
Hence to achieve large reductions in a job's execution time, these techniques need to skip a large proportion of its input data, thus causing large accuracy losses for two reasons. First of all, the entire input data potentially contributes to the job's result accuracy \cite{agarwal2013blinkdb}. Secondly, in the learning process of many ML applications such classification, recommendation, and regression, different data points have different correlations to result accuracy \cite{chakradhar2010best}. For example, in kNN classification and CF-based recommendation applications, the data points having \emph{smaller distances} to test points and \emph{larger weights} (similarities) to test users have higher influences to classification accuracies and prediction errors, respectively. The techniques have so far been focusing on sampling a part of input data to produce approximate results, while mostly ignoring these data points' correlations to result accuracy.

In this paper, we argue that to provide fast job execution while guaranteeing high result accuracy in ML applications, we need to performance approximate processing using the information of the entire input data and to prioritize data processing according to different data points' correlations to result accuracy. To this end, this paper proposes AccurateML, a framework to enable information-aggregation-based approximate processing on MapReduce jobs.
The basic approach taken by AccurateML is to aggregate information of similar input data points to create small aggregated data points in all map tasks of a job with two purposes.
First, AccurateML uses these aggregated data points, which represent an approximation of the entire input data, to produce initial outputs of the map tasks quickly to save the job computation time and to decrease the size of the outputs to reduce the job communication time.
Second, AccurateML uses aggregated data points to identify the parts of input data most related to the job's result accuracy, thus first using these parts to improve the produced outputs in order to minimize result accuracy losses.
Note that the proposed framework is not intended to replace, but rather complement the existing techniques that reduce job execution time based on producing exact results \cite{zaharia2009job,kc2010scheduling,zaharia2010delay,chowdhury2011managing,chen2012joint,verma2012two,zhu2014minimizing,ahmad2014shufflewatcher,huang2015need}.
AccurateML also differs from traditional techniques that pre-compute structures (e.g. samples or wavelets) of input data based on past execution logs and then use these structures to process jobs running on \emph{certain} attributes of input data with accuracy and latency bounds \cite{agarwal2013blinkdb,agarwal2015succinct}. In contrast, AccurateML needs no prior knowledge about the jobs to be processed and it can support jobs running on \emph{arbitrary} attributes of input data.

AccurateML is proposed for ML applications under the assumption the small proportion of critical input data can contributes to majority parts of result accuracy. This assumption is reasonable for many ML applications including classication, regression, recommendation, and clustering applications. To demonstrate the effectiveness of our approach, we have implemented it on Spark \cite{sparkWebSite} and incorporated it with two ML applications, namely a kNN classification application and a CF-based recommendation application.
We evaluated AccurateML using real-world datasets in both applications to study its effectiveness.
The evaluation results show: (i) in each map task, the generation of aggregated data points only takes less than 5\% of the job computation time; (ii) compared to the exact processing results, AccurateML achieves 40.12 times and 31.65 times reduction in the job execution times with result accuracy losses of less than 10\% and 3.5\% in the evaluation of the kNN classification application and the CF-based recommendation application, respectively; (iii) compared to the approach based on producing approximate results, AccurateML reduces the accuracy losses by an average of 2.71 times when using the same job execution times.

The remainder of this paper is organized as follows: Section \ref{MotivatingExamples} discusses several examples to motivate the importance of fast and accurate approximate processing. Section \ref{Section: AccurateML} introduces our approach and Section \ref{Section: Evaluation} evaluates the proposed approach. Section \ref{Section: Related Work} discusses the related work, and finally, Sections \ref{Section: Conclusion} summarizes the work.

\section{Motivating Examples} \label{MotivatingExamples}

To motivate our focus on fast and accurate approximate processing approach, we study the influence of input data on both execution time and result accuracy of MapReduce jobs in the context of ML applications. We analyzed the 25 ML applications in Apache Mahout \cite{mahout} whose core algorithms are classification, clustering, and batch-based CF; and the 35 ML algorithms in Spark Machine Learning Library (MLlib) \cite{mllib} that contains a variety of classification, regression, clustering, and feature reduction applications.

\textbf{Influence of input data on job execution time}. In a typical MapReduce job executing in a Hadoop (Spark) cluster, the input data on which it operates is partitioned into chunks and then allocated to different workers (slave nodes). During job execution, its map tasks process input data and output key-value pairs. These pairs are then transferred to its reduce tasks in the shuffle phase. Subsequently, the reduce tasks compute the final result. Hence the job execution time is determined by its \emph{computation time} of map and reduce tasks, and its \emph{communication time} that is the total data transfer time during input data loading, data shuffling between map and reduce tasks, and result writing. In the following discussions, we focus on the \emph{computation time of map tasks} that usually dominates the job computation time and the \emph{shuffle cost} (i.e. the amount of data transferred in the shuffle phase) that is usually the bottleneck of data transfer.
As shown in Table \ref{table: Categories of ML algorithms in Mahout and MLLib}, we classify the ML applications in Mahout and MLLib into different categories in terms of the two discussed metrics of job execution time and result accuracy.

\begin{table}[h!]
  \caption{Percentages of ML algorithms belonging to different categories}
  \centering
  \begin{tabular}{|p{1cm}|p{3.4cm}|l|l|l|l|}
    \hline
    \multicolumn{2}{|c|}{\textbf{ML library}} &\multicolumn{2}{|c|}{Mahout} & \multicolumn{2}{|c|}{MLLib} \\
    \hline
    \multicolumn{2}{|c|}{\textbf{Category}}& Yes & No & Yes & No  \\
    \hline
    \multirow{6}{1cm}{Job execution time}&Whether map tasks' computation times are proportional to input data size?&\multirow{3}{0.5cm}{96.00}&\multirow{3}{0.5cm}{4.00}&\multirow{3}{0.5cm}{97.14}&\multirow{3}{0.5cm}{2.86}\\
    \cline{2-6}
    &Whether a job's shuffle cost is proportional to input data size?&\multirow{3}{0.5cm}{72.00}&\multirow{3}{0.5cm}{28.00}&\multirow{3}{0.5cm}{42.86}&\multirow{3}{0.5cm}{57.14}\\
    \hline
    \multirow{4}{1cm}{Result accuracy}&Whether a job's result accuracy is influenced by the ratio of the processed input data? &\multirow{4}{0.5cm}{72.00}&\multirow{4}{0.5cm}{28.00}&\multirow{4}{0.5cm}{74.29}&\multirow{4}{0.5cm}{25.71}\\
    \hline
  \end{tabular}
  \label{table: Categories of ML algorithms in Mahout and MLLib}
\end{table}

\emph{Map tasks' computation time}. In over 95\% of the studied ML applications, map tasks' computation times in a MapReduce job are proportional to their input data sizes. In a small percentage of applications such as Stochastic Gradient Descent (SGD) based parameter estimation, the computation time depends on the algorithm's number of iterations, where each iteration only processes one data point.

\emph{Shuffle cost}. In 72.00\% and 42.86\% of the applications (e.g. CF-based recommendation, k-means clustering, and stratified sampling) in Mahout and MLLib, a MapReduce job's shuffle cost is proportional to its input data size. For the remaining applications, the outputs of map tasks are fixed: they are statistics (e.g. likelihood ratios in collocation identification), learned parameters (e.g. variables in linear regression), or discovered frequent patterns (e.g. sequential patterns in FP-growth and association rules). These outputs (that is, the transferred data in the shuffle phase), therefore, are independent of the size of input data.


\textbf{Evaluation of example ML applications}.
As an example of a ML application on MapReduce, we tested the \emph{kNN classification} on a Spark cluster with eight workers, each worker has 2 CPU cores and 2 GB memory.
The tested input data (i.e. the Multiple Features Factor dataset \cite{MultipleFeaturesfac}) has 2.3 million points, the map tasks of a job thus need to calculate the distances between a test point and all these points. These map tasks' outputs are the test points' $k$ closest data points and they are fixed. When classifying 10,000 test points, the job takes 82 minutes to complete.
As a second example, we tested the \emph{CF-based recommendation} using the Netflix Challenge dataset \cite{NetflixDataSet} on the same cluster. The map tasks of this job need to scan about 10 million ratings to make a prediction. When making predictions for about 10,000 ratings, these tasks' output size (i.e. the transferred data volume in the shuffle) is approximately 50 times larger than that (714 MB) of the input data, and the job execution time is about 113 minutes.

\emph{Result accuracy}. We note that in a wide range of ML applications such as classification, recommendation, regression, and clustering applications, restricting the size of the input data influence result accuracy.
As shown in Table \ref{table: Categories of ML algorithms in Mahout and MLLib}, 72.00\% and 74.29\% of the applications in Mahout and MLLib belong to this category. For the remaining applications, they either perform computations over the entire input data (e.g. matrix decomposition algorithms such as Singular Value Decomposition (SVD) and QR decomposition), or only need fixed input data (e.g.  a probability distribution in Markov chain Monte Carlo). These applications are beyond the scope of this work.

\begin{figure}
\centering
\setlength{\abovecaptionskip}{-3pt}
\setlength{\belowcaptionskip}{-15pt}
  \includegraphics[scale=0.53]{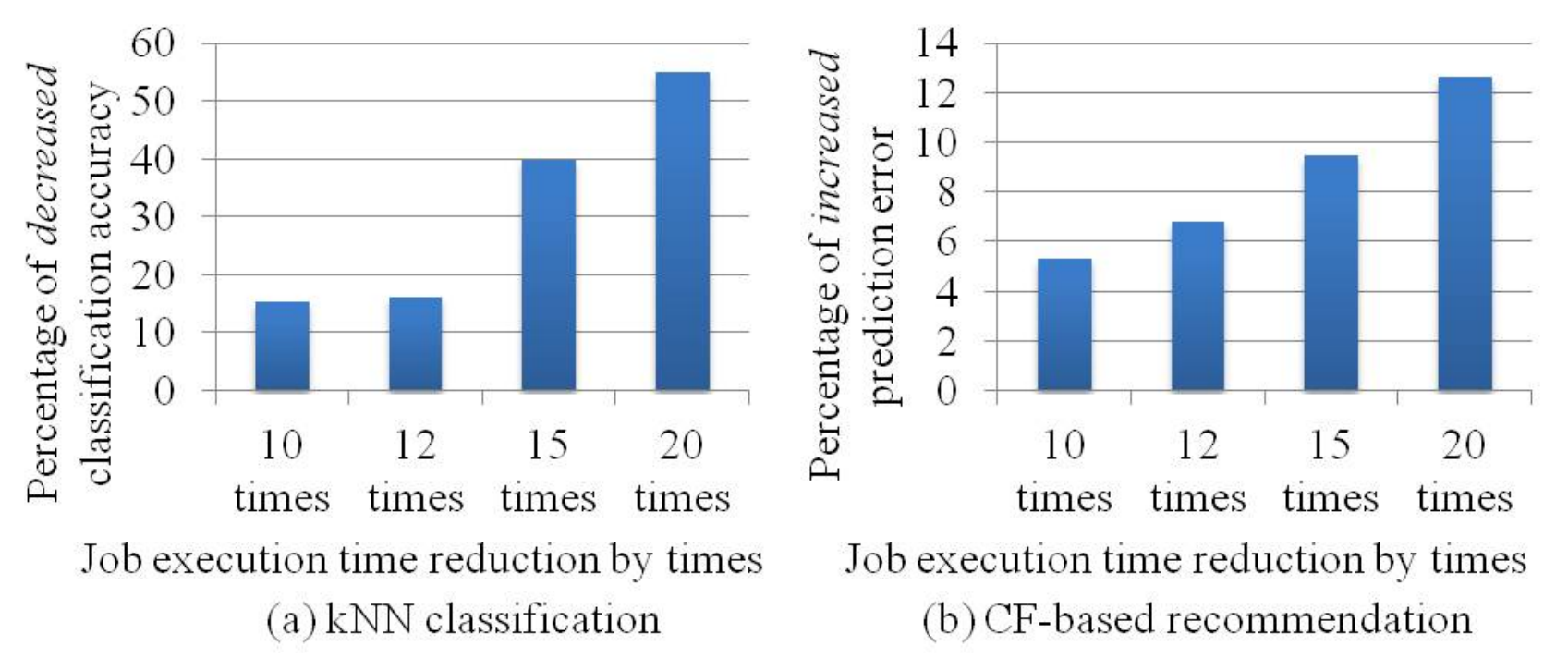}\\
  \caption{The accuracy losses of approximate results when reducing job execution time in two example ML applications}
  \label{Fig: exampleResultAccuracy}
\end{figure}

\textbf{Evaluation of example ML applications}. We tested the kNN classification and the CF-based recommendation applications under the same experimental settings as above. Figure \ref{Fig: exampleResultAccuracy} plots the relationship between the reduced job execution times and the result accuracy losses in the produced approximate results. In kNN classification and CF-based recommendation, the accuracy losses are the percentages of \emph{decreased} classification accuracies and \emph{increased} prediction errors divided by the accuracies and the errors of exact results, respectively.
We can see that when reducing job execution times by 10 to 20 times to achieve fast learning process, existing techniques incur considerable accuracy losses.

\section{AccurateML} \label{Section: AccurateML}

\subsection{Overview} \label{Sec: Overview}

\begin{figure}
\centering
\setlength{\abovecaptionskip}{-3pt}
\setlength{\belowcaptionskip}{-15pt}
  \includegraphics[scale=0.43]{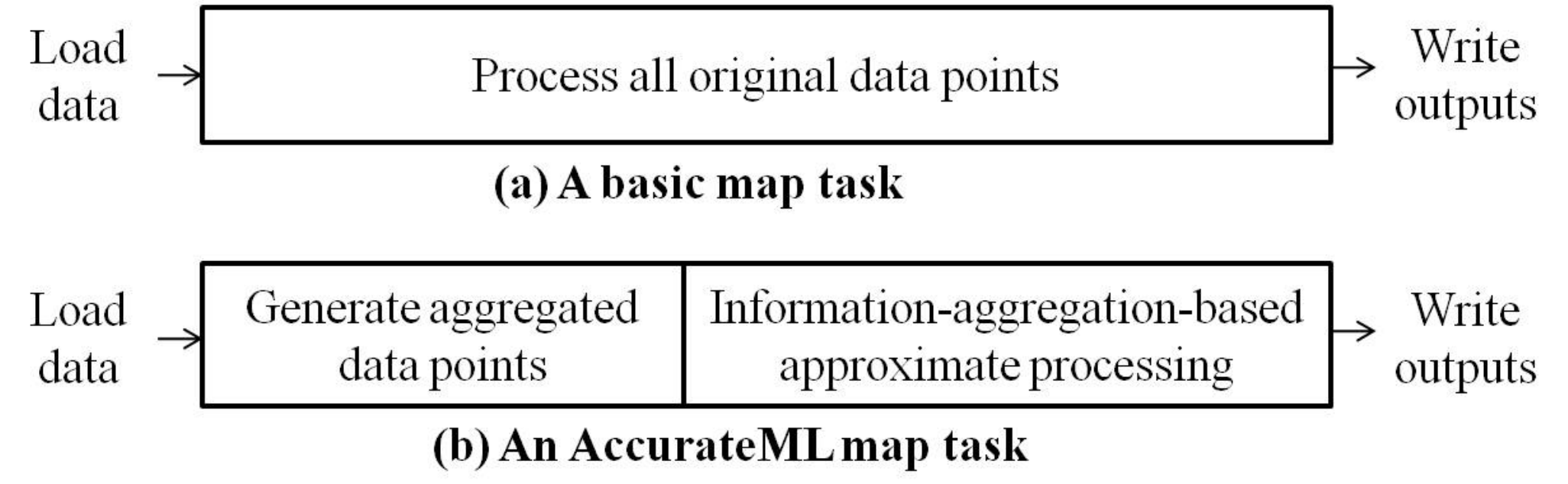}\\
  \caption{Comparison of a basic map task and an AccuracyML map task}
  \label{Fig: Comparison of a basic Map task and a Map task in AccuracyML}
\end{figure}

AccurateML is presented to enable the information-aggregation-based approximate processing in map tasks of MapReduce jobs.
As shown in Figure \ref{Fig: Comparison of a basic Map task and a Map task in AccuracyML}(a), after loading the data, a basic map task produces exact outputs by processing all the original data points in the input data. Its outputs are transferred to reduce tasks in the shuffle phase.
In contrast, AccurateML re-structures the map task's computation step into two steps, as shown in Figure \ref{Fig: Comparison of a basic Map task and a Map task in AccuracyML}(b).

\textbf{Generating aggregated data points}. This step divides the input data into multiple parts using locality sensitive hashing (LSH) \cite{datar2004locality} and then generates several aggregated data points and an index file. Each \emph{aggregated} data point represents the summarized information of a part of similar \emph{original} data points in the input data. The \emph{index file} records the mapping relationship between each aggregated data point and the original data points represented by it.
The detailed generation process is explained in Section \ref{Sec: Generation of aggregated data points}.


\textbf{Information-aggregation-based approximate processing}. Using the aggregated data points, this step produces an approximate output for each map task of a job using two stages. The first stage produces an initial approximate output using the aggregated data points, which are sufficiently small such that the production process can be completed quickly even when handling large input data. By processing the aggregated data points, this stage also estimates the correlations between different parts of the input data and the job's result accuracy. Based on this estimation, the second stage improves the produced output by first processing the most accuracy-related parts of original data points to minimize the result accuracy loss.
Section \ref{Sec: Approximate Processing for Non-iterative ML Applications} explains this step.

\subsection{Generating aggregated data points} \label{Sec: Generation of aggregated data points}

The basic idea of generating aggregated data points is to group similar data points in input data and store their statistical information in aggregated data points to preserve data similarity. In AccurateML, LSH \cite{datar2004locality} is used in grouping similar data points for three reasons:
(1) LSH can map similar data points to the same buckets (that is, the same aggregated data points) with high probability. This is because data points close in the feature space have small distances (e.g. Euclidean distance when $s=2$ in Definition 1). LSH guarantees that the chance of mapping two points $\vec{d}$ and $\vec{d^{\prime}}$ to the same bucket grows as their distance $\parallel \vec{d}-\vec{d^{\prime}}\parallel _s$ decreases (Definition 2);
(2) LSH can control the number of the generated aggregated data points, namely these points' approximation level to the input data, by adjusting the bucket number in mapping. Hence, a larger bucket number means a larger number of aggregated data points and a smaller number of original data points represented by each of them;
(3) LSH works efficiently for large-scale and high dimensional data.
Based on LSH, the generation process has two steps.


\textbf{Definition 1 (Distance measure)}. Given two $n$-dimensional data points $\vec{d}=(x_1, x_2,...,x_n)$ and $\vec{d^{\prime}}=(x^{\prime}_1, x^{\prime}_2,...,x^{\prime}_n)$, the distance between $\vec{d}$ and $\vec{d^{\prime}}$ is calculated as an $l_s$ norm: $\parallel \vec{d}-\vec{d^{\prime}}\parallel _s = \sqrt[s]{\Sigma_{i=1}^n (x_i-x^{\prime}_i)^s}$.

\textbf{Definition 2 (LSH)}. Given two data points $\vec{d}$ and $\vec{d^{\prime}}$, and their distance $\parallel \vec{d}-\vec{d^{\prime}}\parallel _s$, a hash function $h(.)$, which maps a $n$-dimensional data point $\vec{d}$ into a bucket id $h(\vec{d})$, is locality sensitive if it satisfies both conditions that follow:

1. if $\parallel \vec{d}-\vec{d^{\prime}}\parallel _s$ is small (the similarity of $\vec{d}$ and $\vec{d^{\prime}}$ is high), then with high probability $h(\vec{d}) = h(\vec{d^{\prime}})$;

2. if $\parallel \vec{d}-\vec{d^{\prime}}\parallel _s$ is large (the similarity of $\vec{d}$ and $\vec{d^{\prime}}$ is low), then with high probability $h(\vec{d}) \neq h(\vec{d^{\prime}})$;

\emph{Step 1. Grouping similar data points using LSH}. This step operates on the input data and maps similar data points into the same buckets using LSH. A bucket enclosing multiple \emph{original} data points corresponds to an \emph{aggregated} data point. For $l_s$ norm, AccurateML uses a popular LSH hash function as follows \cite{datar2004locality}:
\begin{equation}
h(\vec{d})= \lfloor \frac{\vec{a}\cdot\vec{d}+b}{w}  \rfloor
\label{Equation: LSHFunction}
\end{equation}
where $\vec{a}$ is a $n$-dimensional vector in which each component is drawn from a \emph{p-stable} distribution, $\vec{a}\cdot\vec{d}$ denotes the dot product of $\vec{a}$ and $\vec{d}$, $b$ is uniformly drawn from $[0, w)$ and $w$ is large constant.

In the mapping, this step selects a bucket number to decide the \textbf{compression ratio} (the number of \emph{original} data points divided by the number of \emph{aggregated} data points) such that a sufficient number of aggregated data points are generated to enable the fine-grained differentiation of the input data represented by them.
The number of aggregated data points should also be much smaller (e.g. 10 or 100 times smaller when the compression ratio is 10 or 100) than the number of original input data points to guarantee the short job execution time. This step outputs an index file using the mapping results.


\emph{Step 2. Information aggregation of original data points}. According to the index file, this step obtains each aggregated data point's corresponding \emph{original} data points and aggregates their information. In aggregation, the averages of original data points's feature values are calculated according to Equation \ref{Equation: AggregatedDataPoint}.

\textbf{Definition 3 (Aggregated data points)}. Suppose a $n$-dimensional aggregated data point $\vec{ad}=(y_1, y_2,...,y_n)$ corresponds to a set of $m$ $n$-dimensional original data points \{$\vec{d}^{(1)}$, $\vec{d}^{(2)}$,..., $\vec{d}^{(m)}$\}, where $\vec{d}^{(i)}=(x^{(i)}_1, x^{(i)}_2,...,x^{(i)}_n)$ ($i=1,...,m$), then for $j=1$ to $n$:
\begin{equation}
y_j = \frac{\Sigma_{i=1}^m x^{(i)}_j}{m}
\label{Equation: AggregatedDataPoint}
\end{equation}

\begin{figure}
\centering
\setlength{\abovecaptionskip}{-3pt}
\setlength{\belowcaptionskip}{-15pt}
  \includegraphics[scale=0.43]{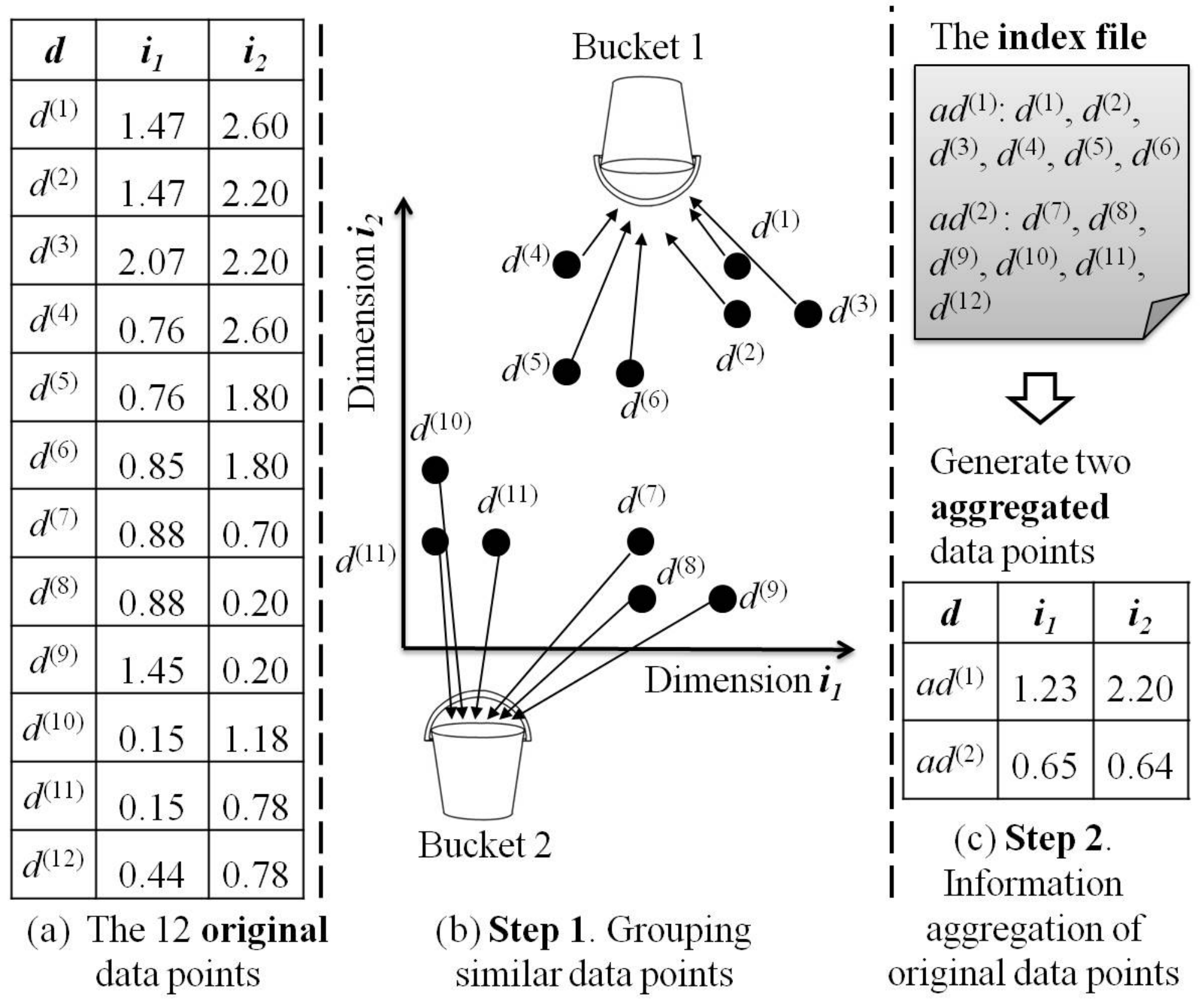}\\
  \caption{An example of LSH-based generation of aggregated data points}
  \label{Fig: An example of LSH-based generation of aggregated data points}
\end{figure}

Figure \ref{Fig: An example of LSH-based generation of aggregated data points} shows an example of generating aggregated data points.
There are 12 original data points in the input data and each point has two features (Figure \ref{Fig: An example of LSH-based generation of aggregated data points} (a)). Step 1 groups similar data points into the same buckets to preserve data similarity (Figure \ref{Fig: An example of LSH-based generation of aggregated data points} (b)). An index file is generated according to the grouping results in the two buckets. Using this file, step 2 generates two aggregated data points $\vec{ad}^{(1)}$ and $\vec{ad}^{(2)}$, each one aggregates information of six original data points.

\subsection{Information-aggregation-based Approximate Processing} \label{Sec: Approximate Processing for Non-iterative ML Applications}

In a map task of a MapReduce job, the steps of information-aggregation-based approximate processing are detailed in Algorithm \ref{AAAPForNonIterativeMLApps}.
An initial output $ao$ of this task is first produced using the aggregated data points (line 1).
During this process, Algorithm \ref{AAAPForNonIterativeMLApps} also estimates the \emph{correlations} (Definition 4) between different parts of the input data and the job's result accuracy to guide the accuracy-aware processing of original data points. This estimation is based on two observations.

\textbf{Definition 4 (Correlation to result accuracy)}. An aggregated (original) data point's correlation to result accuracy, denoted noted by $c$, is the improvement in result accuracy brought by processing this data point.

First, processing an aggregated data point $\vec{ad}^{(i)}$ gives an estimation of the correlation $c_i$ between this point and the result accuracy. For example, in the kNN classification and the CF-based recommendation applications, the correlations are the \emph{negative distance} between an aggregated data point and a test point and the \emph{weight} between an aggregated user and an active user, respectively.

Second, $\vec{ad}^{(i)}$ represents a set $D_i$ of original data points with similar feature values. Hence a higher value of $c_i$ indicates a larger accuracy improvement brought by processing the data points in $D_i$.
For example, in kNN classification, the class label of a test point is decided its $k$ nearest neighbors in the training set.
A smaller distance between $\vec{ad}^{(i)}$ and the test point (that is, a larger $c_i$) indicates the original data points in $D_i$ are closer to the test point. That is, processing these data points has a higher chance of finding the test point's actual $k$ nearest neighbors and thus having a larger probability of classifying it correctly.
Similarily, in CF-based recommendation, result accuracy is measured by the error between the predicted and actual ratings, a higher weight (i.e. a larger $c_i$) means the original users in set $D_i$ have higher similarities to the active user $u$ on average. Processing these users thus has a larger influence on improving $u$'s predicted rating.

Based on the estimated correlations, the algorithm first ranks the aggregated data points (line 2), and then uses the ranking order of each aggregated data point to determine the ranking order of its corresponding set of original data points (line 3). Subsequently, the algorithm sequentially uses the ranked sets to improve output $ao$ (line 4 to 10). The improvement process iteratively executes under the condition that the number $i$ of the processed sets is smaller than or equal to the \textbf{refinement threshold} $\varepsilon_{max}$, which defines the maximal ratio of sets of original data points to be processed in the improvement.
The setting of this threshold saves computations by making the algorithm only to process a proportion of original data points that are most related to the result accuracy.

\begin{algorithm}[htbp]
\caption{Information-aggregation-based approximate processing in a map task}
\label{AAAPForNonIterativeMLApps}
\algsetup{
linenosize=\small,
linenodelimiter=.
}
\begin{algorithmic}[1]
\REQUIRE
\{$\vec{ad}^{(1)}$, $\vec{ad}^{(2)}$,...,$\vec{ad}^{(k)}$\}: the $k$ aggregated data points; \\
$c_i$: $\vec{ad}^{(i)}$'s correlation to the job's result accuracy ($1 \leq i \leq k$);\\
$D_i$: the set of original data points represented by $\vec{ad}^{(i)}$; \\
$ao$: the approximate output of the map task; \\
$\varepsilon_{max}$: the refinement threshold.\\

\STATE Process \{$\vec{ad}^{(1)}$, $\vec{ad}^{(2)}$,...,$\vec{ad}^{(k)}$\} to obtain the initial output $ao$ and the $k$ correlations $c_1$ to $c_k$;
\STATE Rank \{$\vec{ad}^{(1)}$, $\vec{ad}^{(2)}$,...,$\vec{ad}^{(k)}$\} in descending order according to their correlations to result accuracy;
\STATE Obtain the ranked sets \{$D^{\prime}_1$,$D^{\prime}_2$,...,$D^{\prime}_k$\} according to the ranking orders of aggregated data points;
\STATE $i$=0; //$i$ is the index of aggregated data points;
\WHILE{($i \leq k \times \varepsilon_{max}$)}
    \FOR{each original data point $\vec{d} \in D^{\prime}_i$}
        \STATE Process $\vec{d}$ to improve $ao$;
    \ENDFOR
    \STATE $i$=$i+1$;
\ENDWHILE
\STATE Return $ao$.
\end{algorithmic}
\end{algorithm}

\subsection{Implementations} \label{Sec: Implementations}

AccurateML is implemented in Java and its module of \emph{generating aggregated data points} is implemented based on an open source package of LSH \cite{SparkLSH}.
Incorporating its module of \emph{Accuracy-aware approximate processing} into a ML application does not require any modification in the data processing algorithm, but controlling the input data fed to the algorithm. We incorporated AccurateML into two ML applications: a kNN classification application and a CF-based recommendation application.
Both applications are implemented on Spark \cite{sparkWebSite}, a popular MapReduce-like framework built upon Hadoop. We introduce the two applications as follows.

\textbf{The kNN classification application}.
Although conceptually simple, the kNN method is a classic approach \cite{cover1967nearest} that provides a core function of many algorithms in fields such as statistical classification and pattern recognition.
This method also has many features that are common to a wide class of ML algorithms. In the context of classification problems, the basic kNN algorithm classifies a test point $\vec{q}$ by linearly scanning all data points in a training set with known class labels and setting the $k$ ones whose distances are closest to $\vec{q}$ as its $k$ nearest neighbors. The algorithm then assigns $\vec{q}$ to the same class as that of the majority of its nearest neighbors.

\textbf{The CF-based recommendation application}.
The user-based CF algorithm \cite{su2009survey} is a popular data-driven recommendation technique that predicts an active user's rating (preference score) for a target item (product) based on existing ratings from similar users. The input data is a user-item rating matrix that stores the user historical ratings (preference scores) for different items. For an active user $u$, the algorithm predicts $u$'s rating on a target item $i$ using two steps.
The first step calculates the weight (e.g. Pearson's correlation coefficient) $w(u,v)$ between user $u$ and any neighborhood user $v$ who has rated the same item $i$ in the matrix. After calculating the weights, the second step generates the prediction of user $u$'s rating on item $i$ by taking a weighted average of all ratings of item $i$ from user $u$'s neighborhood users: $p(u,i)= \bar{r_{u}}+\frac{\sum_{v \in I} w(u,v) \times (r_{v,i}-\bar{r_v})}{\sum_{v \in I} |w(u,v)|}$,
where $I$ is the set of all users that have rated item $i$, $r_{v,i}$ denotes user $v$'s rating of item $i$, and $\bar{r_u}$ is the average rating of all items rated by user $u$, and $|w(u,v)|$ denotes the absolute value of the weight $w(u,v)$.

\section{Evaluation} \label{Section: Evaluation}

In this section, we first introduce the experimental set-up (Section \ref{Section:Experimental Settings}). AccurateML's salient feature in quickly producing approximate results of small accuracy losses is then demonstrated by comparison with exact results (Section \ref{Section: Evaluation of Trade-off Between Job Execution Time and Result Accuracy}) and approximate results produced using existing techniques (Section \ref{Section:Comparison to the Approximate Processing Approach}).

\subsection{Experimental Settings} \label{Section:Experimental Settings}

\textbf{Hardware Configurations}. The experiments were conducted on a nine-node Spark cluster, in which a 1Gb ethernet network card connects one master node and eight workers. Each node is equipped two Intel Xeon E5645 processor cores, 32 GB of DRAM, and one 1 TB 7200RPM SATA disk drive.

\textbf{Software Environments}. We use the same software configuration for all the workloads and run them in distributed mode. Each worker has two executors. All the cluster nodes run Linux Ubuntu 14.04.1. The KVM, JDK versions are 1.7.91, 1.7.0, respectively. The Hadoop and Spark versions are 2.6.0 and 1.5.2, respectively.

\textbf{Tested workload and dataset}. We test two workloads based on the implementation of AccurateML on two ML applications in Section \ref{Sec: Implementations}.
For the kNN classification workload, the input data is the Multiple Features Factor dataset \cite{MultipleFeaturesfac}, which includes 2.3 million data points belonging to 10 classes and each point has 217 features. From this dataset, we randomly selected about 0.5\% of data points as the test points, and the remaining points form the training set.
For the CF-based recommendation workload, the input data is the Netflix Challenge dataset \cite{NetflixDataSet}, which is a rating matrix with 48,019 users (lines), 17,700 items (columns), and about 10 million ratings.
From this dataset, we randomly selected 1,00 users as the active users. For each active user, 20\% of the items are randomly selected to form the test set, while the remaining 80\% of items and the items from all the other users form the training set.

\textbf{Evaluation metrics}. Both performance and accuracy metrics are used to evaluate the ML applications. The \emph{performance} metric is each job's execution time. The \emph{accuracy} metric is the percentage of accuracy losses, which denotes the percentage of decreased accuracies in approximate results when comparing to accuracies of exact results that are produced using the entire input data.

In classification, the \emph{accuracy} is measured by the prediction accuracy, which denotes the proportion of test points that are correctly classified.
In recommendation, the \emph{accuracy} is measured by the root-mean-square error (RMSE) \cite{su2009survey}, which denotes the errors between the predicted and actual values of ratings. Formally, RMSE is a weighted average error that measures the prediction accuracy for all the target items in a test set $T$: RMSE = $\sqrt{\frac{\sum_{i \in T}(p(u,i)-r_{u,i})^2}{n_T}}$, where $n_T$ represents the number of items in set $T$, $p(u,i)$ is the item $i$'s predicted rating and $r_{u,i}$ is its actual rating.

\subsection{Evaluation of Trade-off Between Job Execution Time and Result Accuracy} \label{Section: Evaluation of Trade-off Between Job Execution Time and Result Accuracy}

The evaluations in this section first show MapReduce jobs' execution times in terms of computation time of map tasks and shuffle cost, and then discuss the trade-off between job execution time and result accuracy.

\textbf{Evaluation settings}. In AccurateML, both job execution time and result accuracy are determined by two parameters: (1) the \emph{compression ratio} that decides the number of aggregated data points used to produce initial outputs of map tasks; and (2) the \emph{refinement threshold} that decides the number of original data points used to refine the outputs.
In the evaluations that follow, three compression ratios (10, 20, and 100) were tested, which means each aggregated data point corresponds to an average of 10, 20, and 100 original data points, respectively; 10 refinement thresholds (0.01 to 0.1) were tested, which means the most accuracy-related 1\% to 10\% of original data points are processed. For each test, the input data of both workloads is divided into 100 partitions for parallel execution. That is, there are 100 map tasks in each job and the average of their evaluation results is report. In addition, the value of $k$ is set to 5 in the kNN classification workload.

\textbf{Evaluation of map tasks' computation time}.
In this evaluation, we divide an AccurateML map task into four parts: grouping similar data points using LSH, information aggregation of original data points, producing initial outputs of map tasks, and refining the outputs by processing original data points. For each part, we report its \emph{percentage computation time}, which denotes the execution time of this part divided by the execution time of a basic map task that processes the entire input data.

\begin{figure*}
\centering
  \includegraphics[scale=0.55]{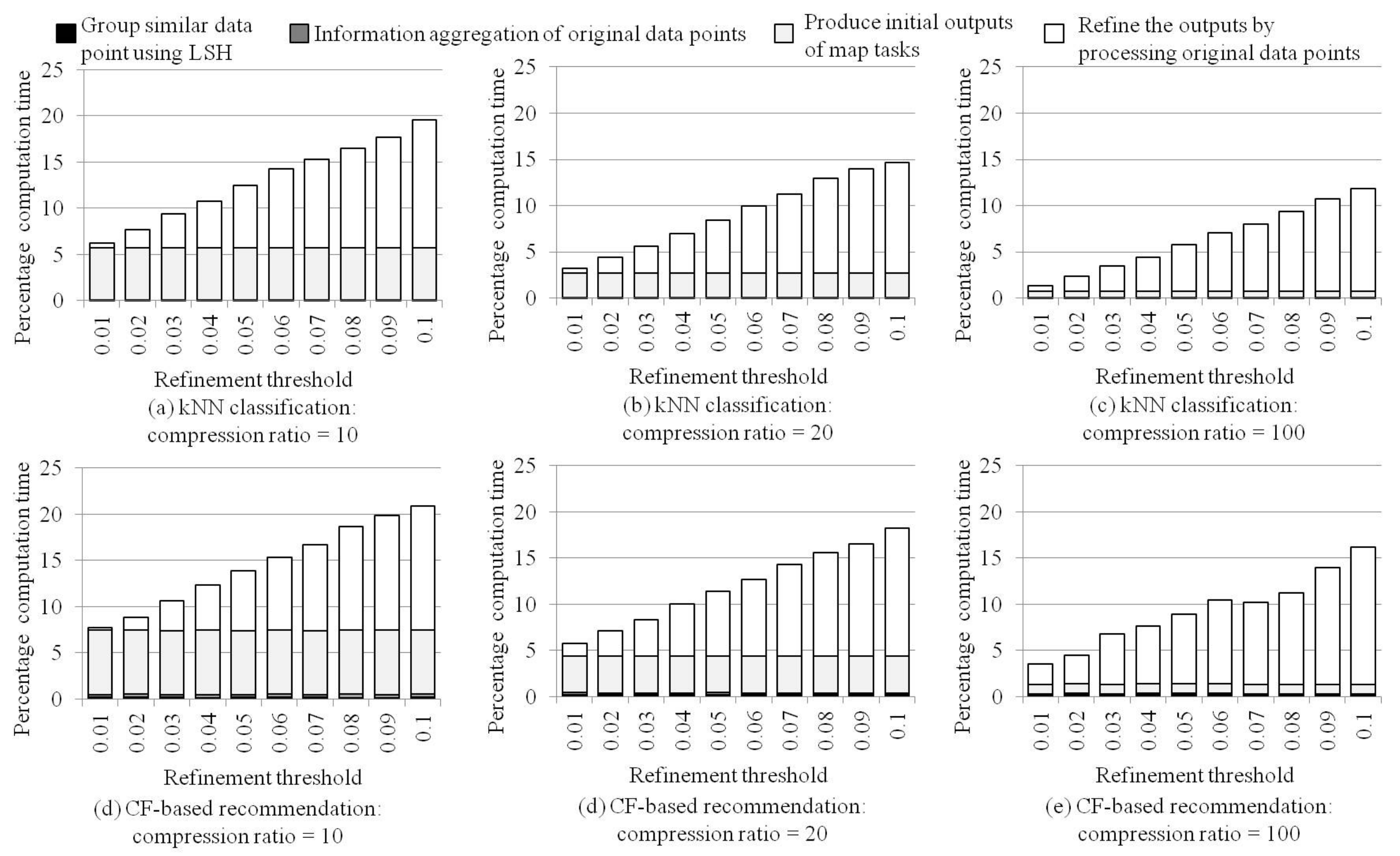}\\
  \caption{Percentage computation time breakdown for map tasks in AccurateML}
  \label{Fig: Percentage computation time breakdown for map tasks in AccurateML}
\end{figure*}

Figure \ref{Fig: Percentage computation time breakdown for map tasks in AccurateML} shows the percentage computation time of the four parts under different values of compression ratios and refinement thresholds. We can see that in all cases, the execution times of the first two parts (i.e. \emph{grouping similar data points using LSH} and \emph{information aggregation of original data points}) are two orders of magnitude smaller than the computation time of the basic map task.
This is because these parts have much lower time complexities compared to the ML algorithm.
Hence in an AccurateML job, a large proportion (more than 95\%) of the computation time is determined by the remaining two parts. Specifically, the percentage computation time of the \emph{producing initial outputs of map tasks} part ranges from 0.65\% to 6.97\%. This percentage is inversely proportional to the compression ratio. That is, a larger compression ratio means shorter times used to produce initial outputs. In the \emph{refining the outputs by processing original data points} part, the percentage computation time ranges from 0.29\% to 14.85\%, which corresponds to the refinement threshold used in approximate processing.
When considering all the four parts, the execution times of AccurateML map tasks are between 1.35\% and 20.90\% of the execution times of the basic map tasks.

\begin{figure}
\centering
  \includegraphics[scale=0.58]{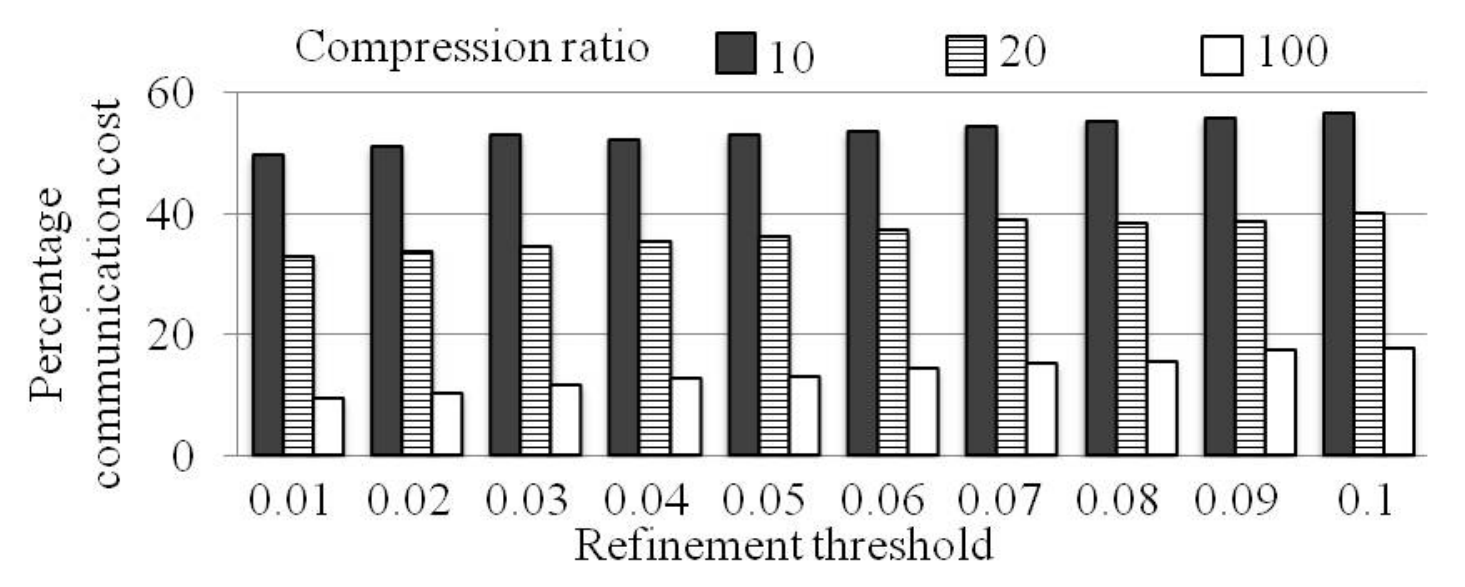}\\
  \caption{Percentage shuffle cost of MapReduce jobs in AccurateML for the CF-based recommendation workload}
  \label{Fig: Percentage communication time of MapReduce jobs in AccurateML for the CF-based recommendation application}
\end{figure}

\textbf{Evaluation of shuffle cost}.
In this evaluation, we report \emph{percentage shuffle cost}, which denotes the transferred data amount in the shuffle phase of an AccurateML job divided by that amount of a basic MapReduce job.
In the kNN classification workload, the outputs of map tasks (i.e. test points' $k$ nearest neighbors) are fixed and thus the job shuffle cost is independent of the input data size. By contrast, in the CF-based recommendation workload, the outputs of map tasks are active users' neighborhood users and the number of these users depends on the input size. Hence AccurateML can reduce the job shuffle costs of this workload. Figure \ref{Fig: Percentage communication time of MapReduce jobs in AccurateML for the CF-based recommendation application} lists each job's percentage shuffle cost. We can see that this percentage ranges from 9.48\% to 56.61\% and it is primarily determined by the compression ratio.

\begin{figure}
\centering
\setlength{\abovecaptionskip}{-3pt}
\setlength{\belowcaptionskip}{-15pt}
  \includegraphics[scale=0.56]{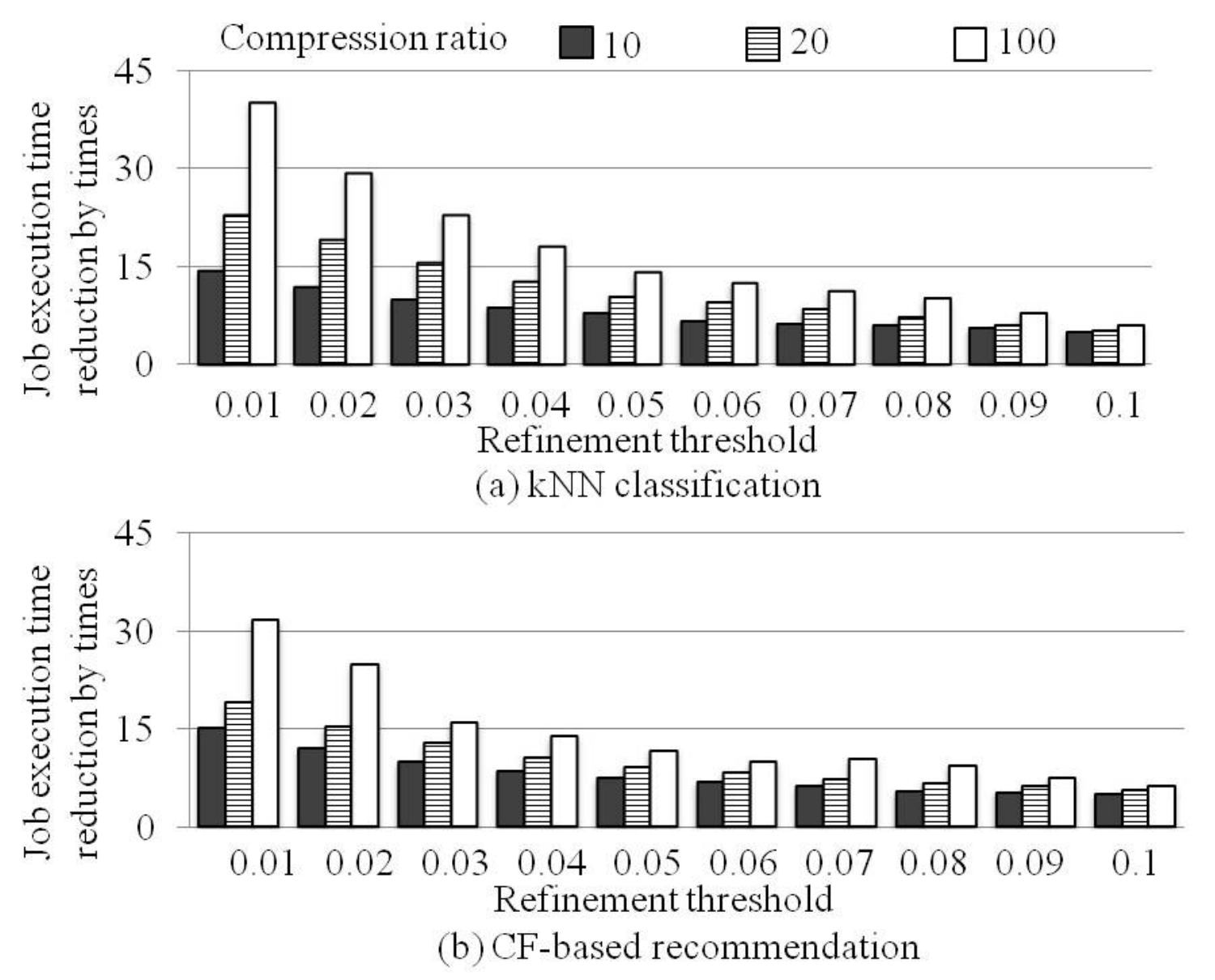}\\
  \caption{Comparison of job execution times between the AccurateML results and the exact results}
  \label{Fig: Reduced total execution time of MapReduce jobs in AccurateML}
\end{figure}

\textbf{Evaluation of job execution time and result accuracy}.
Figure \ref{Fig: Reduced total execution time of MapReduce jobs in AccurateML} shows the job execution time reduction by times when comparing the AccurateML results to the exact results.
We can observe that the jobs produced by AccurateML achieve significant reductions in execution times in all cases. Depending on the values of compression ratios and refinement thresholds, the execution times are reduced by an average of 12.40 times in the kNN classification workload and by an average of 10.85 times in the CF-based recommendation workload.

\begin{figure}
\centering
\setlength{\abovecaptionskip}{-3pt}
\setlength{\belowcaptionskip}{-15pt}
  \includegraphics[scale=0.56]{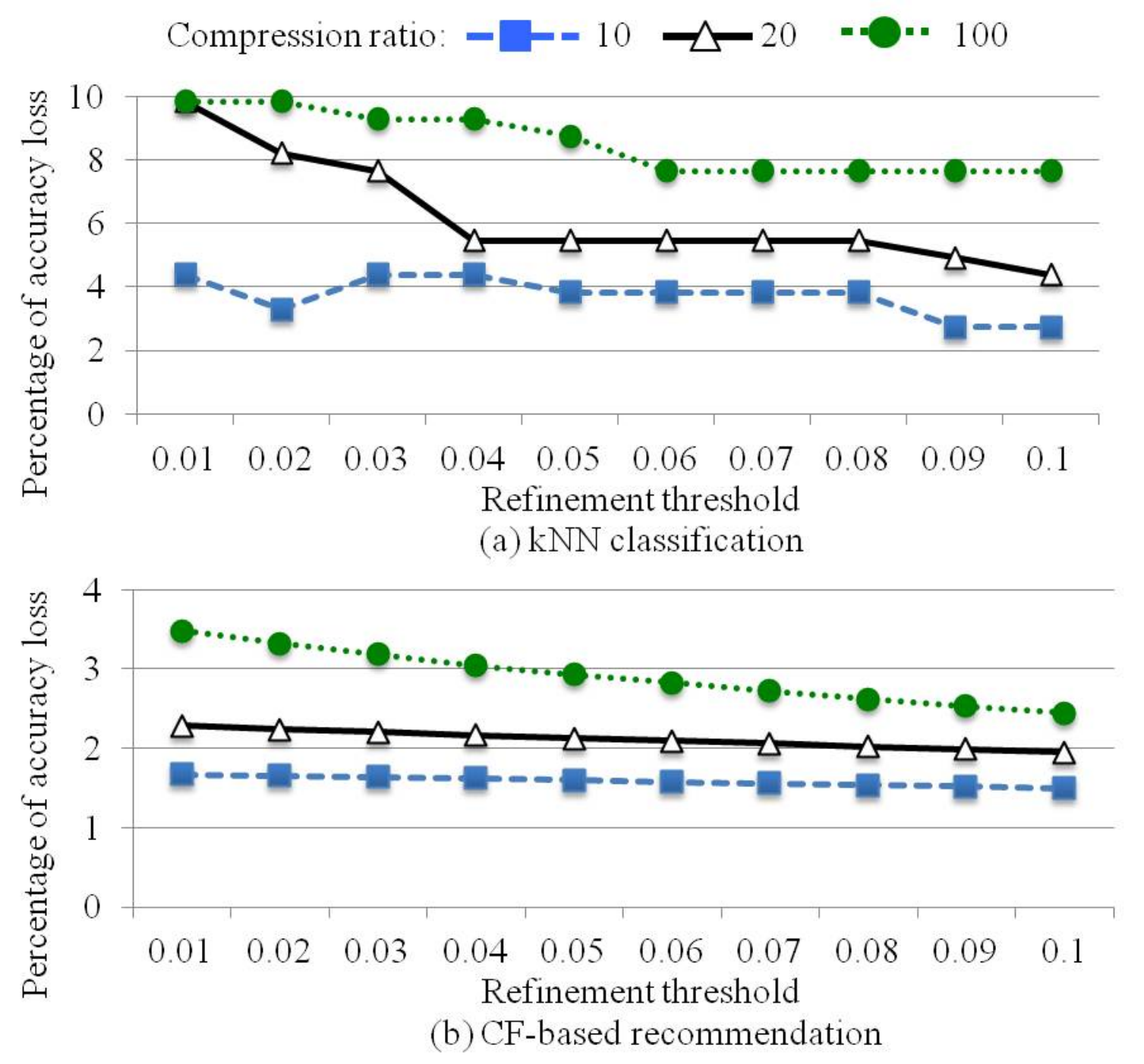}\\
  \caption{Percentages of accuracy losses in the AccurateML results}
  \label{Fig: Accuracy losses of approximate results in AccurateML}
\end{figure}

To achieve such reduction in execution times, Figure \ref{Fig: Accuracy losses of approximate results in AccurateML} shows the AccurateML results' percentages of accuracy loss. We can see that even using short job execution times (that is, setting large compression ratios and small refinement thresholds), the produced results still have small accuracy losses: they are smaller than 10\% and 4\% for the kNN classification workload and the CF-based recommendation workload, respectively. When using a small compression ratio such as 10, these percentages are smaller than 4.37\% and 1.67\% while the job execution time can still be reduced by 5 to 15 times. This is because the aggregated data points used to produce approximate results represent a fine-grained approximation of the entire input data. When processing a small proportion of the most accuracy-related original data points, the accuracy losses can be further reduced.

\textbf{Results}. \emph{Compared to the exact processing approach, AccurateML achieves reductions in job execution time either by 40.12 and 31.65 times with accuracy losses of 9.84\% and 3.48\%, or by 14.30 times and 15.16 times with accuracy losses of 4.37\% and 1.67\% in the evaluations of the kNN classification workload and the CF-based recommendation workload, respectively}.

\subsection{Comparison to the Approximate Processing Approach} \label{Section:Comparison to the Approximate Processing Approach}
Following the evaluation settings of Section \ref{Section: Evaluation of Trade-off Between Job Execution Time and Result Accuracy}, we compare AccurateML with the existing approximate processing approach that reduces job execution times by processing subsets of randomly sampled input data \cite{condie2010mapreduce, pansare2011online,li2011platform,laptev2012early, agarwal2014knowing}.
To make our comparisons fair, the same job execution times are permitted in generating all the approximate results with the two compared approaches.

\begin{figure}
\centering
\setlength{\abovecaptionskip}{-3pt}
\setlength{\belowcaptionskip}{-15pt}
  \includegraphics[scale=0.56]{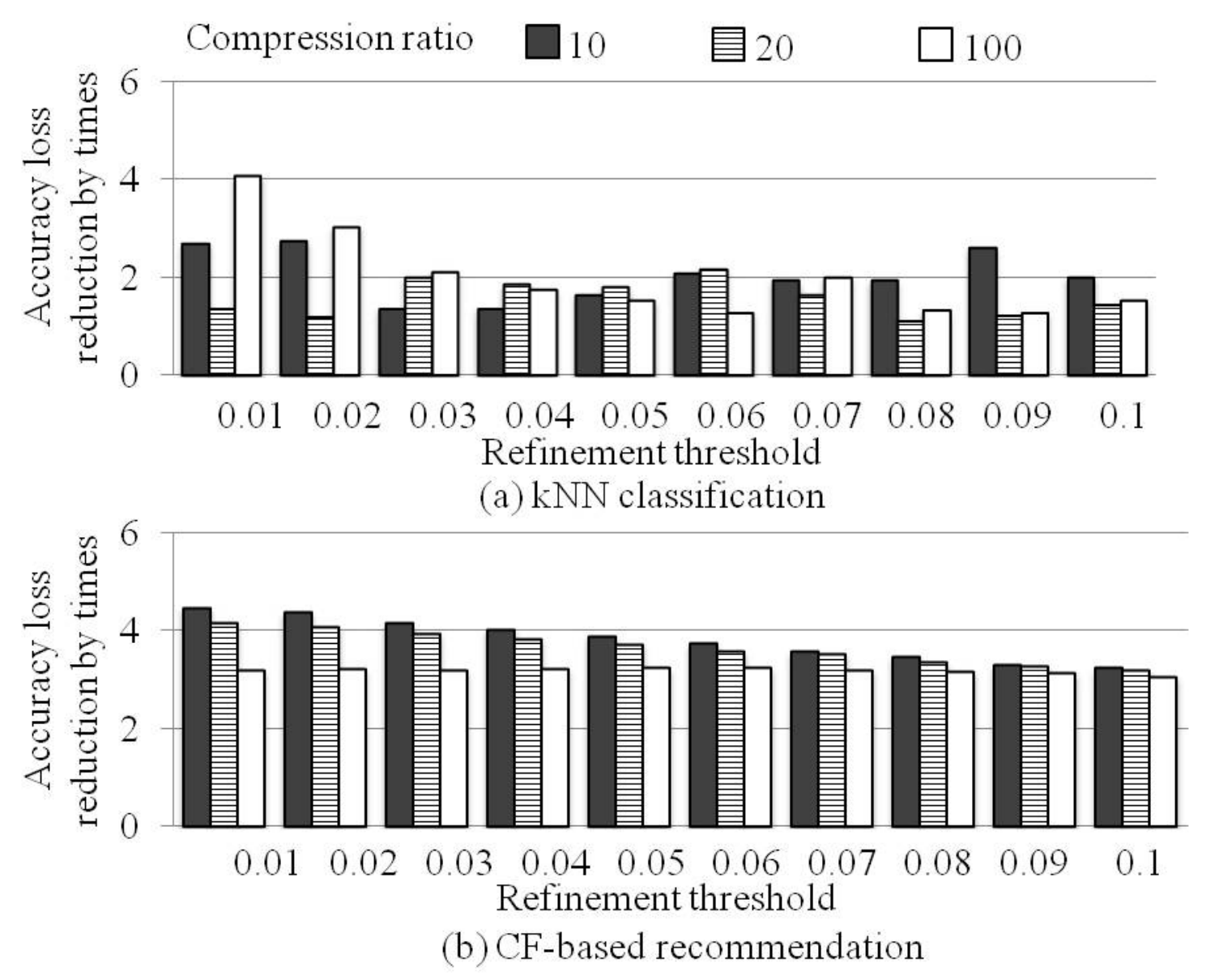}\\
  \caption{Comparison of result accuracy losses between AccurateML and the existing approximate processing approach}
  \label{Fig: Reduced accuracy loss of approximate results in AccurateML}
\end{figure}

\textbf{Evaluation results}. In comparative experiments, 30 approximate results were produced in both approaches. Figure \ref{Fig: Reduced accuracy loss of approximate results in AccurateML} demonstrates the \emph{accuracy loss reduction by times} when comparing the AccurateML results to the approximate results produced by the compared approach. We can see that AccurateML is obviously superior than the compared approach by resulting in much smaller accuracy losses.
This is because to achieve large execution time reductions, a large proportion of information in the input data is either compressed in the aggregated data points (AccurateML) or discarded randomly (the compared approach).
The retained statistical information in the aggregated data points has a higher level of approximation to the entire input data, thus resulting in less accuracy losses.


\begin{figure}
\centering
\setlength{\abovecaptionskip}{-3pt}
\setlength{\belowcaptionskip}{-15pt}
  \includegraphics[scale=0.56]{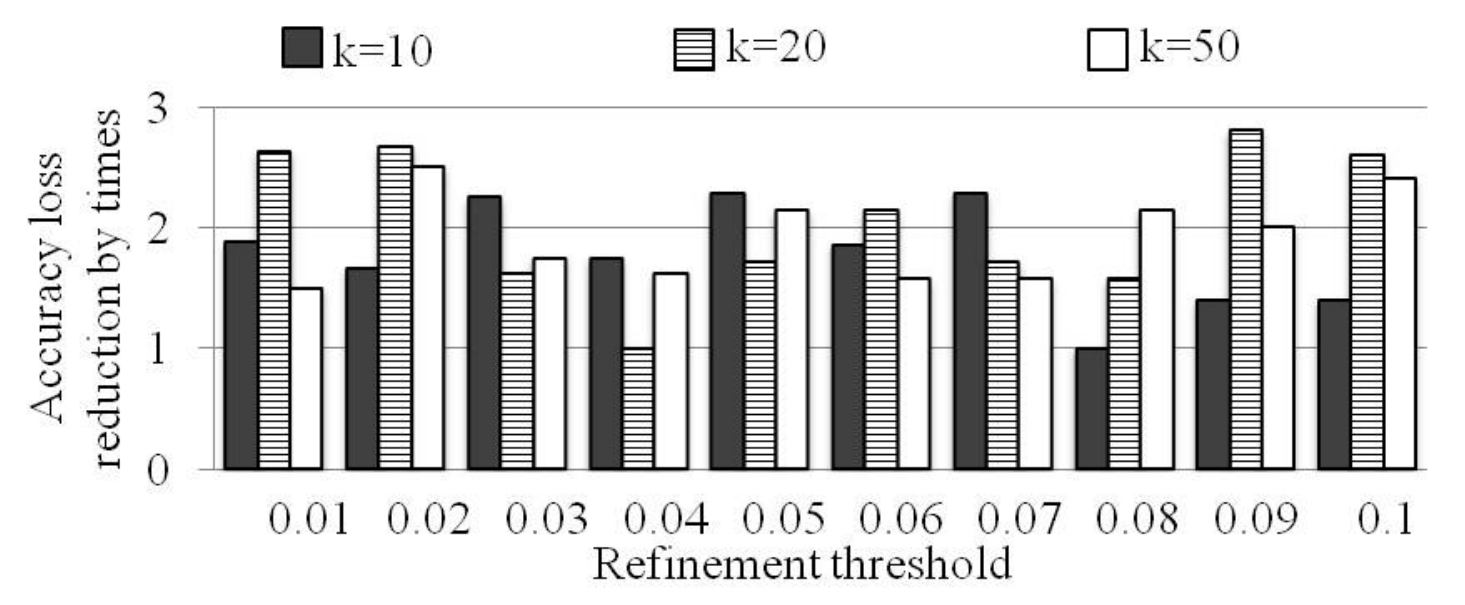}\\
  \caption{Comparison of result accuracy losses between AccurateML and the existing approximate processing approach using different values of $k$ in the kNN classification workload}
  \label{Fig: Reduced accuracy loss of approximate results under different values of $k$ for the CF-based recommendation application}
\end{figure}

\textbf{Discussion of the influence of algorithmic parameters}. We note that in many ML applications, the setting of parameters in their learning algorithms influences result accuracy. Examples include the number of variables used in regression model construction; the number of centroid in clustering; and the number $k$ of nearest neighbors in kNN classification.
We therefore take the kNN classification workload as an example and repeat the above comparative experiment (compression ratio is 10) by testing different values (10, 20, and 50) of $k$.
The evaluation result in Figure \ref{Fig: Reduced accuracy loss of approximate results under different values of $k$ for the CF-based recommendation application} shows that AccurateML consistently achieves much smaller (1.91 times smaller on average) accuracy losses than the compared approach.

\textbf{Results}. \emph{When comparing to the existing approximate processing approach using the same job execution time, AccurateML achieves 1.89 times and 3.55 times reductions in result accuracy losses in the evaluations of the kNN classification workload and the CF-based recommendation workload, respectively}

\section{Related Work} \label{Section: Related Work}

Reducing execution times of MapReduce jobs has attracted much attention in recent years. Many approaches have been proposed based on producing \emph{exact results} and they typically fall into three categories. The first category of approaches dynamically manages the execution orders of multiple MapReduce jobs based on their cost models \cite{kc2010scheduling}, past execution logs \cite{verma2012two}, or utilities of violating deadlines \cite{huang2015need}. The second category of approaches proposes new MapReduce schedulers to address two key problems in job execution \cite{zaharia2009job,tiwari2015classification}: data locality (placing tasks on nodes that having their input data) \cite{zaharia2010delay} and the dependence between map and reduce tasks \cite{chen2012joint,zhu2014minimizing}. The third category of approaches focuses on improving the performance of data transfer (e.g. the shuffle phase) in MapReduce jobs \cite{chowdhury2011managing,ahmad2014shufflewatcher}. Our approximate processing approach forms a complement to the above techniques. In this section, we discuss related work based on producing \emph{approximate results}.

Producing approximate results in ML applications executing on environments with limited time and resources has been extensively studied. We now review two major categories of existing work developed for MapReduce jobs.

\textbf{Sampling based approximate processing}. Online aggregation transforms the traditional batch-oriented processing of MapReduce jobs into an interactive process. That is, it first produces an initial approximate result and then continuously refines it until its estimated accuracy bound (e.g. the accuracy is within a value range with 95\% confidence) reaches a specified one \cite{condie2010mapreduce, pansare2011online, laptev2012early,agarwal2014knowing}. During the computation process, this technique controls the execution time of a MapReduce job by restricting the size of its input data. Hence to achieve fast job execution, this technique skips a large proportion of input data in approximate result production, thus incurring large accuracy losses because all the skipped data potentially contributes to result accuracy. In contrast, AccurateML performs computations over small aggregated data points to produce quick initial results. It also improves the results using the parts of input data most related to result accuracy, thus resulting in high result accuracy while also providing short execution time.

\textbf{Pre-computed structure based approximate processing}. Based on workload characteristic of past logs, some techniques pre-compute specialized structures (e.g. samples, histograms, or wavelets) of input datasets. Each structure can be used to process a MapReduce job running on certain attributes of input data \cite{agarwal2013blinkdb,agarwal2015succinct}. Although these techniques provide both accuracy and execution time bounds for MapReduce jobs, they are impractical to process ML applications that need to process arbitrary attributes of input data. Hence these techniques are orthogonal to AccurateML, which requires no prior knowledge and uses aggregated data points to represent an approximation of the entire input data with all attributes.


\section{Conclusion} \label{Section: Conclusion}

In this paper, we presented AccurateML, an information-aggregation-based approximate processing framework for both fast execution and high result accuracy of ML applications on MapReduce.
AccurateML is based on two key ideas: (1) it aggregates information of input data to create small aggregated data points, thus enabling MapReduce jobs producing initial results quickly despite handling large input data;
(2) it estimates the correlations between different parts of the input data and the jobs' result accuracies using the aggregated data points, thus minimizing accuracy losses by first using the most accuracy-related input data to improve the results.
Evaluation results using real workloads and datasets demonstrate the effectiveness of AccurateML at bringing considerable reductions in job execution times while only causing small accuracy losses.

\section*{Acknowledgments}
We thank the anonymous reviewers for careful review of our paper. This work is supported by the National Natural Science Foundation of China (Grant No. 61502451) and the National Key Research and Development Plan of China (Grant No. 2016YFB1000601). Rui Han is the corresponding author.

\bibliographystyle{plain}
\bibliography{references}

\end{document}